\documentclass[12pt]{article}
\title{\bf
Exact Solutions of Linearized Schwinger-Dyson Equation of Fermion
Self-Energy
}
\author{
{\bf 
Bang-Rong Zhou\thanks{Electronic mailing address: zhoubr@sun.ihep.ac.cn}} \\
\normalsize China Center of Advanced Science and Technology \\
\normalsize ( World Laboratory ) P.O.Box 8730, Beijing 100080,
{\bf China}\\
{\normalsize and}  \\
\normalsize Department of Physics, Graduate School, Academia Sinica, Beijing
100039, {\bf China\thanks{Mailing address.}}
}
\date {}
\begin{document}
\hoffset = -1 truecm
\voffset = -2 truecm
\baselineskip = 12pt
\maketitle

\begin{abstract}
The Schwinger-Dyson equation of fermion self-energy in the linearization
approximation is solved exactly in a theory with gauge and effective
four-fermion interactions. Different expressions for the indepedent
solutions which respectively submit to irregular and regular ultraviolet 
boundary condition are derived and expounded. \\  \\
\indent PACS number(s): 12.38.Aw, 12.38.Bx, 02.30.Hq, 02.30.Gp

(Running title: EXACT SOLUTIONS OF LINEARIZED S-D EQUATION...) 

\end {abstract}

\newpage

\noindent {\Large \bf  I. Introduction} \\
\indent Owing to non-linearity of the Schwinger-Dyson (S-D)
${\rm equation}^{1-7}$ of the fermion self-energy, together with running
of the gauge coupling
constant being included in, it is scarcely possible to obtain an analytic
solution of the equation.  As a result, one usually has to solve it by numerical
${\rm method}^{8-10}$.
However, one could still get some analytic solutions of the S-D equation if 
some linearization approximation to the equation is made. The resulting 
analytic solutions will be quite useful for discussions of chiral symmetry 
breaking.  In this paper, we will derive and expound the general exact 
analytic solutions of the S-D equation of fermion self-energy in the 
linearization approximation. \\
\indent Consider a theory with vectorial gauge and chirally-invariant effective
four-fermion interactions. When running of the gauge coupling constant
is taken into account, the S-D equation of the fermion self-energy $\Sigma(p^2)$
in the Landau gauge, after Wick rotation and angular integration, will have the
following ${\rm form}^{11}$:

$$\Sigma(p^2) = m_0(\Lambda) + \frac{3 C_2(R)}{16 {\pi}^2}\int_{0}^{{\Lambda}^2}
dk^2\frac{k^2\Sigma(k^2)}{k^2+{\Sigma}^2(k^2)}
\frac{\bar{g}^2({\rm max}(p^2,k^2))}{{\rm max}(p^2,k^2)} +
\frac{hd(R)}{2{\pi}^2}\int_{0}^{{\Lambda}^2}dk^2\frac{k^2\Sigma(k^2)}
{k^2+{\Sigma}^2(k^2)}\eqno(1.1)$$
\noindent $m_0$ is the bare fermion mass, $\Lambda$ is the ultraviolet (UV)
momentum cut-off, $C_2(R)$ is the eigenvalue of the squared Casimir operator of
the "color" gauge group in the representation $R$ of the fermion field $\psi$
with the dimension $d(R)$, $h$ is the strength of the chirally-invariant
four-fermion interaction $[{(\bar{\psi}\psi)}^2-{(\bar{\psi}\gamma_5\psi)}^2]$
and $\bar{g}^2({\rm max}(p^2,k^2))$ is the conventional approximation of the
running gauge coupling constant $\bar{g}^2((p-k)^2)$ defined by
$$\bar{g}^2({\rm max}(p^2,k^2)) = \left\{
                         \begin{array}{ll}
                         \bar{g}^2(p^2)  &{\rm\ if} \  \  p^2>k^2 \\
                         \bar{g}^2(k^2)  &{\rm\ if} \  \  k^2>p^2
                         \end{array}
                   \right.\eqno(1.2)$$
\noindent Set
$$x \equiv p^2, \ \ y \equiv k^2\eqno(1.3)$$
\noindent Eq.(1.1) will be reduced to that
$$\Sigma(x) = m_0 + \frac{3C_2(R)}{16{\pi}^2\beta_0}\left[\frac{1}{x\tau(x)}
\int_0^xdy\frac{y\Sigma(y)}{y+{\Sigma}^2(y)}+\int_x^{{\Lambda}^2}dy\frac{
\Sigma(y)}{\tau(y)(y+{\Sigma}^2(y))}\right]+\frac{hd(R)}{2{\pi}^2}
\int_0^{{\Lambda}^2}dy\frac{y\Sigma(y)}{y+{\Sigma}^2(y)} \eqno(1.4)$$
\noindent In Eq.(1.4) we have used a continuous ${\rm Ansatz}^{6}$ of the running
gauge coupling constant
$$\bar{g}^2(q^2) = 1/\beta_0\tau(q^2)\eqno(1.5)$$
\noindent where
$$\tau(q^2) = \ln(\frac{q^2}{\mu^2}+\xi)\eqno(1.6)$$
\noindent and
$$\beta_0=\left[11C_2(G)-\sum_f4T(R_f)N_f\right]/48\pi^2\eqno(1.7)$$
\noindent with the standard denotations in gauge theory.  In the flavor sum 
$\sum_f$ in Eq.(1.7), besides the fermions corresponding to
$\psi$, all the lighter colored fermions will also be included in. The scale
parameter $\mu$ is optional and the parameter $\xi$ is required to be greater
than 1 so as to avoid the infrared (IR) singularity of $\bar{g}^2(q^2)$. \\
\indent The integral equation (1.4) is equivalent to the following differential
equation
$$\omega(x)\Sigma''(x)+[\omega'(x)+1]\Sigma'(x)=-\frac{b}{\tau(x)}
\frac{\Sigma(x)}{[x+\Sigma^2(x)]}\eqno(1.8)$$
\noindent together with the IR boundary condition
$$\Sigma'(0)=-\frac{b}{2(\ln{\xi})\Sigma(0)}\eqno(1.9)$$
\noindent and the UV boundary condition
$${\left\{\left[1+\frac{a}{b}x\tau(x)\right]\omega(x)\Sigma'(x)+
\Sigma(x)\right\}}_{x={\Lambda}^2}=m_0(\Lambda)\eqno(1.10)$$
\noindent where
$$b=\frac{3C_2(R)}{16\pi^2\beta_0},\ \ \ \ a=\frac{hd(R)}{2 \pi^2}\eqno(1.11)$$
\noindent and
$$\omega(x)={\left[\frac{1}{x}+\frac{1}{(x+\xi \mu^2)\tau(x)}\right]}^{-1}
\eqno(1.12)$$
\noindent We emphasize the following points: 1) The non-linearity of Eq.(1.8)
is embodied in the term in the right-handed side of the equation.  2) The IR
boundary condition (1.9) comes from the $x \rightarrow 0$ limit of the equation
$$\Sigma'(x)=b\frac{d}{dx}\left(\frac{1}{x\tau(x)}\right)\int_0^xdy
\frac{y\Sigma(y)}{y+\Sigma^2(y)}\eqno(1.13)$$
\noindent which is a result of Eq.(1.4). Since Eq.(1.13) is valid for any $x$,
we can define the IR boundary condition by Eq.(1.13) at some non-zero value of
$x$, instead of at $x=0$, if $\Sigma'(x)$ is calculable at that value of $x$.
This fact will have a close bearing on the actual solution of the following
linearized S-D equation.  3) The term in Eq.(1.4) relevant to the four-fermion
interactions contains no $x$ hence appears only in the UV boundary condition
(1.10) rather than in Eq.(1.8) itself. \\
\indent In Sect.II we will state the linearization approximation for Eq.(1.8)
and a necessary change of the IR boundary condition of the solution.  In 
Sect.III the derivation of the general analytic solutions of the linearized 
S-D equation will be given in detail and in Sect.IV we will conclude with some 
discussions on forms and features of the independent solutions. \\
\noindent {\Large \bf II. The linearization approximation}  \\
\indent The non-linear S-D equation (1.4) or (1.8) has no analytic solution.
Some analytic solutions could be obtained merely in the linearization
approximation of the equation.  To see how to make this approximation, we first
consider the IR and UV asymptotic solutions of Eq.(1.8).  \\
\indent In the region where $x$ is small, Eq.(1.8) becomes
$$x\Sigma''(x)+2\Sigma'(x)+
\frac{b}{(\ln\xi)}\frac{\Sigma(x)}{[x+{\Sigma}^2(x)]}=0 \eqno(2.1)$$
\noindent Suppose $\Sigma(x)$ has the form
$$\Sigma(x) = x^s(a_0+a_1x+a_2x^2+\cdots),  \ \ \ \ {\rm when} \  x \ {\rm is \
small}
\eqno(2.2)$$
\noindent and substituting Eq.(2.2) into the IR boundary condition (1.9) i.e.
$$\lim \limits_{x\to 0}\Sigma'(x)\Sigma(x)=-\frac{b}{2 \ln\xi}\eqno(2.3)$$
\noindent we obtain that
$$\lim \limits_{x\to 0}\left\{a_0^2sx^{2s-1}+(2s+1)a_0a_1x^{2s}+
\left[(s+1)a_1^2+2(s+1)a_0a_2\right]x^{2s+1}+\cdots\right\} = 
-\frac{b}{2 \ln\xi}\eqno(2.4)$$
\noindent The only possibility to satisfy Eq.(2.4) with a real $a_0$ is to set
$s=0$.  With this result, substituting Eq.(2.2) into Eq.(2.1) we will have
$$2a_1+6a_2x+12a_3x^2+\cdots+\frac{b}{(\ln \xi) a_0^2}\left[
a_0-\left(\frac{1}{a_0}+a_1\right)x+
\left(\frac{1}{a_0^3}+3\frac{a_1}{a_0^2}+
\frac{a_1^2}{a_0}-a_2\right)x^2+\cdots\right] = 0 \eqno(2.5)$$
\noindent Thus we may express $a_1, a_2,a_3 \cdots$ by means of $a_0\equiv
\Sigma(0)$ and write down the solution of Eq.(2.1) in small $x$ region which
is consistent with the IR boundary condition (1.9) as follows:
$$\Sigma(x)=\Sigma(0)\left\{1-\frac{b}{2(\ln\xi) {\Sigma}^2(0)}x+
\frac{b}{6(\ln\xi) {\Sigma}^4(0)}\left(1-\frac{b}{2 \ln\xi}\right)x^2\right.$$
$$\left.-
\frac{b}{12 (\ln\xi) {\Sigma}^6(0)}\left[1-\frac{5b}{3\ln \xi}+
\frac{b^2}{3{(\ln\xi)}^2}\right]x^3+\cdots\right\}\eqno(2.6)$$
\noindent where $\Sigma(0)$ is a finite constant. \\
\indent In the region where $x\rightarrow \infty$, we may have two possible
assumptions: 1) ${\Sigma}^2(x)\geq x$ and 2) $\Sigma^2(x)< x$.  If
${\Sigma}^2(x)\geq x $ is assumed, the asymptotic form of Eq.(1.8) will become
that
$$x\left(1-\frac{1}{\ln x}\right)\Sigma''(x)+
\left(2-\frac{1}{\ln x}\right)\Sigma'(x)
=-\frac{b'}{(\ln x)\Sigma(x)}, \ \ b'=\frac{b}{1+\lambda} \ \ {\rm with} \ \
\lambda \leq 1 \eqno(2.7)$$
\noindent where the ratio $x/\Sigma^2(x)$ at $x\rightarrow \infty$ in the 
denominator of the right-handed side of Eq.(2.7) has been replaced 
approximately by the constant $\lambda$. This replacement will not change the 
essential behavior of the asympototic solution. 
Eq.(2.7) remains to be a non-linear equation.  The exact form of its
asymptotic solution is in general unknown.  However, we can always suppose a
typical form of the solution, such as the form of conventional power function, 
so as to  examine whether the assumption that $\Sigma^2(x) \geq x$ when 
$x\rightarrow \infty$ could be consistent with Eq.(2.7) or not.  Thus we set 
that 
$$\Sigma(x)\sim C(\ln x)^rx^s\eqno(2.8)$$
\noindent where $C$ is a constant, then Eq.(2.7) may be reduced to
$$C\left\{-r(r-1){(\ln x)}^{r-3}+r(r-2s-1){(\ln x)}^{r-2}+
[r(2s+1)-s^2]{(\ln x)}^{r-1}+s(s+1){(\ln x)}^r\right\}x^{s-1} $$
$$=-\frac{b'}{C}{(\ln x)}^{-r-1}x^{-s} \eqno(2.9)$$
\noindent When $s\neq 0,-1$, the leading term in the left-handed side is
$Cs(s+1){(\ln x)}^rx^{s-1}$ whose equality to the right-handed side requires
that
$$r=-r-1, \ s-1=-s, \ Cs(s+1)=-b'/C \eqno(2.10)$$
\noindent and they give that $r=-1/2, \ s=1/2$ and $C^2=-4b'/3$.  The results
demand ${\Sigma}^2(x)\sim x/\ln x$, contradictory to the assumption
${\Sigma}^2(x)\geq x$.  When $s=0$ or $-1$, the leading term in the
left-handed side is $C[r(2s+1)-s^2]{(\ln x)}^{r-1}x^{s-1}$ whose equality to
the right-handed side requires that
$$r-1=-r-1, \ s-1=-s, \ C[r(2s+1)-s^2]=-b'/C\eqno(2.11)$$
\noindent and they give that $r=0,\ s=1/2$ and $C^2=4b'$. The result $s=1/2$ is
obviously opposite to the presupposition $s=0$ or $-1$.  Therefore, the
assumption ${\Sigma}^2(x)\geq x $ when $x\rightarrow \infty$ can not be
consistent with Eq.(2.7), at least this is true for the supposed 
asymptotic  form (2.8) of the solution.  Alternatively, we may assume that 
${\Sigma}^2(x)< x$ when $x\rightarrow \infty$.  In this case, Eq.(1.8) 
approximately becomes that
$$\left(1-\frac{1}{\ln x}\right)x{\Sigma}''(x)+\left(2-\frac{1}{\ln x}\right)
\Sigma'(x)=-\frac{b}{(\ln x)}\frac{\Sigma(x)}{x}\eqno(2.12)$$
\noindent Noting that Eq.(2.12) has now been linearized.  We may still
substitute the trial solution (2.8) into Eq.(2.12) and obtain the algebraic
equation
$$[r(2s+1)-s^2]{(\ln x)}^{r-1}x^{s-1}+s(s+1){(\ln x)}^rx^{s-1}=
-b{(\ln x)}^{r-1}x^{s-1}\eqno(2.13)$$
\noindent where the non-leading terms such as ${(\ln x)}^{r-2}, \
{(\ln x)}^{r-3}$ etc. have been neglected.  Eq.(2.13) could be satisfied
if the term with ${(\ln x)}^rx^{s-1}$ is removed by setting
$$s(s+1)=0\eqno(2.14)$$
\noindent i.e. $s=0$ or $s=-1$.  Thus we will obtain the UV asymptotic
solution of $\Sigma(x)$ with the form
$$\Sigma(x)=A{\left(\ln{\frac{x}{\mu^2}}\right)}^{-b}+
   \frac{B}{x/\mu^2}{\left(\ln{\frac{x}{\mu^2}}\right)}^{b-1}\eqno(2.15)$$
\noindent The fact that the UV asymptotic form (2.12) of Eq.(1.8) has been
linearized indicates that the non-linearity of Eq.(1.8) is important only in the
IR region.  Actually, the linearization approximation of Eq.(1.8) is a good one
not only in the UV region but also in the mediate momentum region where $x$ is
not very small.  The approximation can be made by replacing ${\Sigma}^2(x)$ in
the denorminator of the non-linear term in Eq.(1.8) by ${\Sigma}^2(0)$ and this
will result in the following linearized version of Eq.(1.8)
$$\omega(x)\Sigma''(x)+[\omega'(x)+1]\Sigma'(x)=-\frac{b}{\tau(x)}
\frac{\Sigma(x)}{[x+{\Sigma}^2(0)]}\eqno(2.16)$$
\noindent Generally, the ${\Sigma}^2(0)$ in the right-handed side of Eq.(2.16)
may also be replaced by some constant which will be viewed as the average value
of $\Sigma(x)$ over small $x$ region.  So the linearized Eq.(2.16) will be
qualitatively correct for small $x$ where non-linearity is important. \\
\indent To make it be easy to solve Eq.(2.16) analytically we take two further
assumptions.  One of them is to suppose that
$${\Sigma}^2(0)=\xi {\mu}^2\eqno(2.17)$$
This is permissive since both the IR parameter $\xi$ and the scale parameter
$\mu$ are undeterminated theoretically.  Of couse, the constraint that $\xi >1$
will demand that ${\Sigma}^2(0)>{\mu}^2$.  Another assumption is that the
coefficient $\omega(x)$ [Eq.(1.12)] in Eq.(2.16) is replaced approximately by
$$\omega(x)\simeq \frac{\tau(x)}{1+\tau(x)}(x+\xi \mu^2)\eqno(2.18)$$
\noindent This approximation is valid if
$$\frac{x}{\xi \mu^2}\left[1+\frac{1}{\ln{(\frac{x}{\mu^2}+\xi)}}\right] \gg 1
\eqno(2.19)$$
\noindent The condition (2.19) could impose some constraints on $\xi$.  Let
$x/\xi \mu^2\geq r$ then Eq.(2.19) will become that
$$\frac{1}{\ln[(r+1)\xi]} \gg \frac{1-r}{r}\eqno(2.20)$$
\noindent The inequality (2.20) is obviously valid for $r\geq 1$ and it may be
changed into that
$$\xi \ll \frac{1}{r+1}e^{\frac{r}{1-r}} \ \ \ {\rm for} \  \  0<r<1
\eqno(2.21)$$
\noindent Hence, in the case with $r<1$, $\xi$ will have an upper bound.
The upper bound will decrease as $r$ goes down.   Although for some values of
$r$ we could obtain the constraint on $\xi$ which is allowed by phenomenenlonogy
(for example, if $r=3/4$ then the constraint will be $1<\xi \ll 11.477$),in the
following we prefer $r=1$ to $r<1$, i.e. we will use the assumption (2.18)
continuously down to the scale $x=\xi \mu^2$ without the need to consider any
upper bound of $\xi$.  \\
\indent Under the assumptions (2.17) and (2.18), the linearized equation (2.16)
may be reduced to the following form that
$$\frac{\tau(x)}{1+\tau(x)}(x+\xi \mu^2)\Sigma''(x)+
\left\{2-\frac{\tau(x)}{{[1+\tau(x)]}^2}\right\}\Sigma'(x)=
-b\frac{\Sigma(x)}{\tau(x)(x+\xi \mu^2)}\eqno(2.22)$$
\noindent The solution of Eq.(2.22) must submit to the UV boundary condition
(1.10) and some IR boundary condition as well.  Since Eq.(2.22) is inapplicable
in the region where $x<\xi \mu^2$, we will not be able directly to use the IR
boundary condition (1.9) at $x\rightarrow 0$.  Instead, as was indicated in
Sect.I, we may now use Eq.(1.13) and the asymptotic solution (2.6) of
$\Sigma(x)$ in small x region to give a new IR boundary condition at
$x=\xi \mu^2$, i.e.
$${\left.\Sigma'(x)\right|}_{x=\xi {\mu}^2}=
b\frac{d}{dx}{\left.\left(\frac{1}{x\tau (x)}\right)\right|}_{x=\xi {\mu}^2}
\int_0^{\xi {\mu}^2}dy\frac{y\Sigma (y)}{y+{\Sigma }^2(y)} \eqno(2.23)$$
\noindent where we have also made the assumption that the solution (2.6)
of $\Sigma(x)$ in small $x$ region will be approximately extended to
$x=\xi {\mu}^2$.  The use of the IR boundary condition (2.23) instead of
Eq.(1.9) will allow us to be able both to solve Eq.(2.22) exactly and
analytically and to include partially the IR non-linearity of Eq.(2.10) in. \\
\indent It is noted that the scales of physical chiral symmetry breaking will
generally be in the region where $x>\xi {\mu}^2$, so it is reasonable to use
Eq.(2.22) to discuss such kind of problem.  The assumptions that the
approximation (2.18) and the solution (2.6) are extended to $x=\xi {\mu}^2$
respectively from above and from below will at most numerically affect the IR
boundary condition (2.23) of Eq.(2.22) and not produce an essential impact
on physical conclusions. \\
\noindent {\Large \bf III. Solutions of the linearized equation}  \\
\indent We will find out the exact indepedent solutions of the linearized
equation (2.22).  Instead of $x$, let $\tau$ to be the new variable, then
Eq.(2.22) may be changed into that
$$\frac{\tau}{1+\tau}\Sigma''(\tau)+
\left[1+\frac{1}{{(1+\tau)}^2}\right]\Sigma'(\tau)+
b\frac{\Sigma(\tau)}{\tau}=0 \eqno(3.1)$$
\noindent Assume the exact solution of Eq.(3.1) have the form
$$\Sigma(\tau)=e^{s\tau}{\tau}^rf(\tau) \eqno(3.2)$$
\noindent where $s$ and $r$ are constants to be determined.  Substituting
Eq.(3.2) into Eq.(3.1) and dividing each term of the equation by
$e^{s\tau}\tau^{r+1}$, we will obtain the differential equation satisfied
by $f(\tau)$ as follows:
$$(\tau +1)f''(\tau)+[(2s+1)\tau+2(r+s+1)+2(r+1)/\tau]f'(\tau)$$   $$+
\{s(s+1)\tau+[2s(r+1)+s^2+r+b]+[r(r+1)+2s(r+1)+2b]/\tau+
[r(r+1)+b]/{\tau}^2\}f(\tau) = 0 \eqno(3.3)$$
\noindent The UV ($\tau\rightarrow \infty$) asymptotic solutions of Eq.(3.1)
may be obtained from Eq.(3.3) by taking $f(\tau)\equiv {\rm constant}$ and
setting the first and the second term in the coefficient of $f(\tau)$ to be
equal to zeroes as well as neglecting all the terms with the orders $1/\tau$
and above.  The results are denoted by
$$\Sigma_{\rm irreg}^{\rm UV}(\tau)=\tau^{-b} \ \ {\rm and}  \  \
\Sigma_{\rm reg}^{\rm UV}=e^{-\tau}\tau^{b-1} \eqno(3.4)$$
\noindent Eq.(3.4) is obviously consistent with the UV asymptotic form (2.15)
of $\Sigma(x)$.  In order to find out the exact solutions of Eq.(3.1) with the
UV asymptotic forms (3.4), we must set the first and the fourth term in the
coefficient of $f(\tau)$ in Eq.(3.3) to be equal to zeroes, i.e.
$$s(s+1)=0 \eqno(3.5)$$
\noindent and
$$r(r+1)+b=0 \eqno(3.6)$$
\noindent Eq.(3.5) will make $\Sigma(x)$ have the correct UV asymptotic form
(3.4) and Eq.(3.6), whose left-handed side is just the coefficient of
$1/{\tau}^2$, can also be obtained from the small $\tau$ limit of Eq.(3.1) by
setting $\Sigma(\tau)\sim {\tau}^r$, hence it certainly contains the small
$\tau$ behavier of Eq.(3.1).  Since the $r$ given by Eq.(3.6) takes complex
values, the corresponding solutions of $\Sigma(\tau)$ will include complex
constants.  However, a physical solution may be a real linear combination of
the two real components of a complex $\Sigma(\tau)$. \\
\indent Eq.(3.5) gives that $s=0$ and $s=-1$.  We will discuss the two cases
respectively. \\
\indent 1) s=0. In this case, the solution (3.2) becomes $\Sigma^{(0)}(\tau)=
\tau^rf^{(0)}(\tau)$. Then from Eq.(3.3), the equation of $f^{(0)}(\tau)$ can 
be changed into that
$$(z^2-z){f^{(0)}}''(z)-[z^2-\gamma (z-1)]{f^{(0)}}'(z)-(\alpha z-b)f^{(0)}(z)
=0 \eqno(3.7)$$
\noindent where
$$z=-\tau \eqno(3.8)$$
$$\gamma=2(r+1) \eqno(3.9)$$
$$\alpha=b+r \eqno(3.10)$$
\noindent Eq.(3.7) could be solved by means of confluent hypergeometric 
(Kummer) ${\rm function}^{12}$. Noting that if $|z|\rightarrow \infty$, then 
Eq.(3.7) becomes
$$z{f^{(0)}}''(z)+(\gamma -z){f^{(0)}}'(z)-\alpha f^{(0)}(z) = 0 \eqno(3.11)$$
\noindent which is just the confluent hypergeometric equation with the solution
$f^{(0)}= {_{1}F_{1}}(\alpha; \gamma; z)$.  Hence, the solution of Eq.(3.7) may 
be written to be that
$$f^{(0)}(z)=a \ {_{1}F_{1}}(\alpha; \gamma; z)+g(z) \eqno(3.12)$$
\noindent where $a$ is a constant.  Substituting Eq.(3.12) into Eq.(3.7) and
considering that ${_{1}F_{1}}(\alpha; \gamma; z)$ obeys Eq.(3.11) we obtain that
$$(z^2-z)g''-[z^2+\gamma (1-z)]g'+(b-\alpha z)g =
az \  {_{1}F_{1}}'(\alpha;\gamma;z)-a(b-\alpha ) \ 
{_{1}F_{1}}(\alpha;\gamma;z) 
\eqno(3.13)$$
\noindent It is seen that the left-handed sides of both Eq.(3.13) and Eq.(3.7)
have the same form.  In view of ${_{1}F_{1}}(\alpha;\gamma;z)$ has been used in
$f^{(0)}(z)$ we may assume that
$$g(z)=cz \ {_{1}F_{1}}'(\alpha;\gamma;z) \eqno(3.14)$$
\noindent where $c$ is a constant.  Substituting Eq.(3.14) into Eq.(3.13), we
will meet the terms with ${_{1}F_{1}}''(\alpha;\gamma;z)$ and 
${_{1}F_{1}}'''(\alpha;\gamma;z)$.
However, they can be changed into some combinations of the terms with only
${_{1}F_{1}}(\alpha;\gamma;z)$ and ${_{1}F_{1}}'(\alpha;\gamma;z)$ by means of 
Eq.(3.11).  In this way, Eq.(3.13) will be reduced to that
$$c\left(\frac{\gamma}{2}-1\right)z \ {_{1}F_{1}}'(\alpha;\gamma;z)-
c\alpha \ {_{1}F_{1}}(\alpha;\gamma;z) = az \ {_{1}F_{1}}'(\alpha;\gamma;z)-
a(b-\alpha) \ {_{1}F_{1}}(\alpha;\gamma;z)$$
\noindent which demands that
$$a-\left(\frac{\gamma}{2}-1\right)c = 0, $$
$$-(b-\alpha)a+\alpha c = 0 \eqno(3.15)$$
\noindent The $a$ and $c$ have non-zero solution if and only if
$$\alpha-\left(\frac{\gamma}{2}-1\right)(b-\alpha) = 0 \eqno(3.16)$$
\noindent which is obviously satisfied based on Eqs.(3.9), (3.10) and (3.6).
As a result, we have $a=\frac{\gamma}{2}-1=r$ if $c=1$ is taken.  From this
we will have the solution of Eq.(3.7) with the form
$$f^{(0)}(z) = \left(\frac{\gamma}{2}-1\right) \ {_{1}F_{1}}(\alpha;\gamma;z) +
z\frac{d}{dz} {_{1}F_{1}}(\alpha;\gamma;z) \eqno(3.17)$$
\noindent and the corresponding
$$\Sigma^{(0)}(\tau)={\tau}^r\left[\left(\frac{\gamma}{2}-1\right) \
{_{1}F_{1}}(\alpha;\gamma;-\tau)-\frac{\alpha}{\gamma}\tau \  
{_{1}F_{1}}(\alpha+1; \gamma+1;-\tau)
\right] \eqno(3.18)$$
\noindent Depending on the UV asymptotic form, the physically-relevant real 
solutions will have the following two kinds of forms:
$$\left.\matrix{
  \Sigma^{(0)}_{\rm irreg}(\tau) \cr
  \Sigma^{(0)}_{\rm reg}(\tau)  \cr}\right\}=
  \left.\matrix{
  A_i^{(0)}  \cr
  A_r^{(0)} \cr}\right\}\tau^{-\frac{1}{2}-i\eta}
  \left[\left(\frac{\gamma}{2}-1\right) \ {_{1}F_{1}}(\alpha;\gamma;-\tau)-
  \frac{\alpha}{\gamma}\tau \ {_{1}F_{1}}(\alpha+1;\gamma+1;-\tau)\right]+c.c.
  \eqno(3.19)$$
\noindent where one of two roots of the algebraic equation (3.6) of $r$
$$r=-\frac{1}{2}-i\eta, \ \eta=\sqrt{b-\frac{1}{4}} \eqno(3.20)$$
\noindent has been definitely chosen (chosing the other root $r=-\frac{1}{2}+
i\eta$ has no difference because the physical solutions depend on only the
two real components of $\Sigma^{(0)}(\tau)$).  The complex constants $A_i^{(0)}$
and $A_r^{(0)}$ will be determined respectively by the UV asymptotic conditions
$$\Sigma^{(0)}_{\rm irreg}(\tau)\stackrel{\tau \rightarrow \infty}{\rightarrow}
{\Sigma^{\rm UV}_{\rm irreg}(\tau)={\tau}^{-b}} \eqno(3.21)$$
\noindent and
$$\Sigma^{(0)}_{\rm reg}(\tau)\stackrel{\tau \rightarrow \infty}{\rightarrow}
{\Sigma^{\rm UV}_{\rm reg}(\tau)=e^{-\tau}{\tau}^{b-1}} \eqno(3.22)$$
\noindent By using the asymptotic formula of confluent hypergeometric
${\rm function}^{12}$
$${_{1}F_{1}}(\alpha;\gamma; -\tau)
\stackrel{\tau \rightarrow \infty}\rightarrow
\frac{\Gamma(\gamma)}{\Gamma(\alpha)}e^{-\tau}{(-\tau)}^{\alpha -\gamma }+
\frac{\Gamma(\gamma)}{\Gamma(\gamma -\alpha)}{\tau}^{-\alpha } \eqno(3.23)$$
\noindent we can write

$$\left.\matrix{
        \Sigma^{(0)}_{\rm irreg}(\tau) \cr
        \Sigma^{(0)}_{\rm reg}(\tau) \cr }
  \right\}\stackrel{\tau \rightarrow \infty}\rightarrow
  -b\left[\left.\matrix{A_i^{(0)} \cr
                        A_r^{(0)} \cr}\right\}
      \frac{\Gamma(\gamma)}{\Gamma(\gamma -\alpha)} + c.c.\right]{\tau}^{-b}
  +\left[\left.\matrix{A_i^{(0)} \cr
                       A_r^{(0)} \cr}\right\}
      \frac{\Gamma(\gamma)}{\Gamma(\alpha)}{(-1)}^{\alpha-\gamma+1}+c.c.\right]
      e^{-\tau}{\tau}^{b-1} \eqno(3.24)$$
\noindent The conditions (3.21) and (3.22) will lead to that
$$A_i^{(0)}=i(-1)^{\alpha}\frac{\pi}
{b{\rm sinh}(2 \pi \eta){|\Gamma(\alpha)|}^2}
\frac{\Gamma(\alpha)}{\Gamma(\gamma)} \eqno(3.25)$$
\noindent and
$$A_r^{(0)}=i\frac{\Gamma(\alpha)}{\Gamma(\gamma)}
\frac{{\rm sin}(\pi \alpha)}{{\rm sinh}(2\pi \eta)}
\eqno(3.26)$$
\indent 2) s=-1.  In this case the solution (3.2) becomes $\Sigma^{(-1)}(\tau)=
e^{-\tau}{\tau}^rf^{(-1)}(\tau)$ and from Eq.(3.3), $f^{(-1)}(\tau)$ submits to
the equation
$$\tau (\tau +1){f^{(-1)}}''(\tau)-
\left[{\tau}^2-(\gamma -2)\tau -\gamma \right]{f^{(-1)}}'(\tau)-
(\bar{\alpha}\tau + \gamma -b)f^{(-1)}(\tau)=0 \eqno(3.27)$$
\noindent where
$$\bar{\alpha}=1+r-b \eqno(3.28)$$
\noindent The solution $f^{(-1)}(\tau)$ can be expressed by 
${_{1}F_{1}}(\bar{\alpha};\gamma; \tau)$.  Assume that
$$f^{(-1)}(\tau)=\bar{a} \ {_{1}F_{1}}(\bar{\alpha};\gamma;\tau) +
\bar{c}\tau \ {_{1}F_{1}}'(\bar{\alpha};\gamma;\tau) \eqno(3.29)$$
\noindent Substituting Eq.(3.29) into Eq.(3.27) and considering that
${_{1}F_{1}}(\bar{\alpha};\gamma;\tau)$ satisfies the equation
$$\tau \ {_{1}F_{1}}''(\bar{\alpha};\gamma;\tau)+
(\gamma -\tau ) \ {_{1}F_{1}}'(\bar{\alpha};\gamma;\tau)-
\bar{\alpha} \  {_{1}F_{1}}(\bar{\alpha};\gamma;\tau) = 0 \eqno(3.30)$$
\noindent we obtain that
$$[-\bar{a}+(b+\bar{\alpha})\bar{c}]\tau \ 
{_{1}F_{1}}'(\bar{\alpha};\gamma;\tau)+
  [(\bar{\alpha}+b-\gamma )\bar{a}+\bar{\alpha}\bar{c}] \
  {_{1}F_{1}}(\bar{\alpha};\gamma;\tau) = 0 $$
\noindent which demands that
$$\matrix{-\bar{a} &+ &(b+\bar{\alpha})\bar{c} &= 0 \cr
          (\bar{\alpha}+b-\gamma )\bar{a} &+ &\bar{\alpha}\bar{c} &= 0 \cr}
          \eqno(3.31)$$
\noindent The condition on which $\bar{a}$ and $\bar{c}$ have non-zero solution
$$\bar{\alpha}+(b+\bar{\alpha})(b+\bar{\alpha}-\gamma) =0 \eqno(3.32)$$
\noindent is certainly satisfied in view of Eqs.(3.28),(3.6) and (3.9).  Hence
if  taking $\bar{c}=1$ then we will have $\bar{a}=b+\bar{\alpha}=r+1=\gamma /2$
and
$$f^{(-1)}(\tau)=\frac{\gamma}{2} \ {_{1}F_{1}}(\bar{\alpha};\gamma;\tau)+
\tau \frac{d}{d\tau} \ {_{1}F_{1}}(\bar{\alpha};\gamma;\tau) \eqno(3.33)$$
\noindent and correspondinly,
$$\Sigma^{(-1)}(\tau)=e^{-\tau}{\tau}^r\left[\frac{\gamma}{2} \
{_{1}F_{1}}(\bar{\alpha};\gamma;\tau)+
\frac{\bar{\alpha}}{\gamma}\tau \ {_{1}F_{1}}(\bar{\alpha} +1;\gamma +1;\tau)
\right] \eqno(3.34)$$
\noindent The real solutions $\Sigma^{(-1)}_{\rm irreg}(\tau)$  and
$\Sigma^{(-1)}_{\rm reg}(\tau)$ can be expressed by

$$\left.\matrix{\Sigma^{(-1)}_{\rm irreg}(\tau) \cr
                \Sigma^{(-1)}_{\rm reg}(\tau) \cr}\right\}=
  \left.\matrix{A_i^{(-1)} \cr
                A_r^{(-1)} \cr}\right\}e^{-\tau}{\tau}^{-\frac{1}{2}-i\eta}
 \left[\frac{\gamma}{2} \ {_{1}F_{1}}(\bar{\alpha};\gamma;\tau)+
       \frac{\bar{\alpha}}{\gamma}\tau \ 
{_{1}F_{1}}(\bar{\alpha}+1;\gamma+1;\tau)
 \right]+c.c. \eqno(3.35)$$

\noindent where the complex constants $A_i^{(-1)}$ and $A_r^{(-1)}$ will be
determined respectively by the UV asymptotic conditions
$$\Sigma^{(-1)}_{\rm irreg}(\tau)\stackrel{\tau \rightarrow \infty}\rightarrow
\Sigma^{\rm UV}_{\rm irreg}(\tau)={\tau}^{-b} \eqno(3.36)$$
\noindent and
$$\Sigma^{(-1)}_{\rm reg}(\tau)\stackrel{\tau \rightarrow \infty}\rightarrow
\Sigma^{\rm UV}_{\rm reg}(\tau)=e^{-\tau}{\tau}^{b-1} \eqno(3.37)$$
\noindent By using the asymptotic formula of confluent hypergeometric function
$${_{1}F_{1}}(\bar{\alpha};\gamma;\tau)
\stackrel{\tau \rightarrow \infty}\rightarrow
\frac{\Gamma(\gamma)}{\Gamma(\bar{\alpha})}e^{\tau}{\tau}^{\bar{\alpha}-\gamma}
+\frac{\Gamma(\gamma)}{\Gamma(\gamma-\bar{\alpha})}
e^{-i\pi \bar{\alpha}}{\tau}^{-\bar{\alpha}} \eqno(3.38)$$
\noindent we get from Eqs.(3.36) and (3.37) that
$$A_i^{(-1)}=i(-1)^{\alpha}\frac{\Gamma(\bar{\alpha})}{\Gamma(\gamma)}
\frac{{\rm sin}(\pi \bar{\alpha})}{{\rm sinh}(2\pi \eta)} \eqno(3.39)$$
\noindent and
$$A_r^{(-1)}=-i\frac{\pi}{b \ {\rm sinh}(2\pi \eta)|\Gamma(\bar{\alpha})|^2}
  \frac{\Gamma(\bar{\alpha})}{\Gamma(\gamma)}. \eqno(3.40)$$
\noindent {\Large \bf IV. Some discussions of solutions}  \\
\indent We will show that among the four real solutions 
$\Sigma^{(s)}_{\rm irreg}$ and $\Sigma^{(s)}_{\rm reg} \ (s=0,-1)$ derived 
above, only two ones are linearly
independent, consistent with the general conclusion that a second order
differential equation has two independent solutions. \\
\noindent In fact, $\Sigma^{(0)}(\tau)$ and $\Sigma^{(-1)}(\tau)$ differ by only
a constant.  Denote
$$\Sigma^{(0)}(\tau)={\tau}^rf^{(0)}(\tau) \eqno(4.1)$$
\noindent then $f^{(0)}$ will obey Eq.(3.7) or
$$\tau(\tau+1){f^{(0)}}''(\tau)+\left[{\tau}^2+\gamma (\tau +1)\right]
{f^{(0)}}'(\tau) + (\alpha \tau +b)f^{(0)}(\tau) = 0 \eqno(4.2)$$
\noindent Now write
$$f^{(0)}(\tau)=e^{-\tau}h(\tau) \eqno(4.3)$$
\noindent then it is easy to verify that $h(\tau)$ obey the equation
$$\tau(\tau+1)h''(\tau)-\left[{\tau}^2-(\gamma -2)\tau -\gamma \right]h'(\tau)-
(\bar{\alpha}\tau +\gamma-b)h(\tau) = 0 \eqno(4.4)$$
\noindent which has the same form as Eq.(3.27) of $f^{(-1)}(\tau)$.  It follows
from this that $h(\tau)\propto f^{(-1)}(\tau)$.  Hence
$$\Sigma^{(0)}(\tau)=e^{-\tau}{\tau}^rh(\tau)\propto
e^{-\tau}{\tau}^rf^{(-1)}(\tau)=\Sigma^{(-1)}(\tau) \eqno(4.5)$$
\noindent i.e. $\Sigma^{(0)}(\tau)$ and $\Sigma^{(-1)}(\tau)$ differ at most
by a constant and both are linearly dependent. \\
\indent This conclusion may also be reached directly from the explicit
expressions of $\Sigma^{(0)}(\tau)$ and $\Sigma^{(-1)}(\tau)$.  By means of
the property of confluent hypergeometric ${\rm function}^{12}$
$${_{1}F_{1}}(\alpha;\gamma;-\tau)=e^{-\tau}  \ 
{_{1}F_{1}}(\gamma-\alpha;\gamma;\tau)
\eqno(4.6)$$

\noindent we can express $\Sigma^{(0)}(\tau)$ in Eq.(3.18) by

$$\Sigma^{(0)}(\tau)=e^{-\tau}{\tau}^r\left[\left(\frac{\gamma}{2}-1\right)
{_{1}F_{1}}(\gamma -\alpha;\gamma;\tau)-\frac{\alpha}{\gamma}\tau \ 
{_{1}F_{1}}(\gamma -\alpha;\gamma +1;\tau)\right] \eqno(4.7)$$
\noindent From Eqs.(3.9), (3.10) and (3.28) we see that
$$\gamma -\alpha = \bar{\alpha} +1 \eqno(4.8)$$
\noindent By Eq.(4.8) and the recurrence ${\rm relation}^{12}$
$$\gamma \ {_{1}F_{1}}(\bar{\alpha}+1;\gamma;\tau) =
\tau \ {_{1}F_{1}}(\bar{\alpha}+1;\gamma+1;\tau)+
\gamma \ {_{1}F_{1}}(\bar{\alpha};\gamma;\tau) \eqno(4.9)$$
\noindent we may rewrite Eq.(4.7) in the form that

$$\begin{array}{l}\Sigma^{(0)}(\tau)\\
=e^{-\tau}{\tau}^r\left(\frac{\gamma}{2}-1\right)\left[
{_{1}F_{1}}(\bar{\alpha};\gamma;\tau)+\frac{1}{\gamma}
\left(1-\frac{\alpha}{\gamma/2-1}\right)\tau \
{_{1}F_{1}}(\bar{\alpha}+1;\gamma+1;\tau)\right] \\
=e^{-\tau}{\tau}^r\frac{\gamma-2}{2}\left[
{_{1}F_{1}}(\bar{\alpha};\gamma;\tau)+
\frac{2}{\gamma}\frac{\bar{\alpha}}{\gamma}\tau \
{_{1}F_{1}}(\bar{\alpha}+1;\gamma+1;\tau)\right] \\
=\frac{\gamma-2}{\gamma}e^{-\tau}{\tau}^r\left[
\frac{\gamma}{2} \ {_{1}F_{1}}(\bar{\alpha};\gamma;\tau)+
\frac{\bar{\alpha}}{\gamma}\tau \ 
{_{1}F_{1}}(\bar{\alpha}+1;\gamma +1;\tau)\right] \\
=\frac{\gamma -2}{\gamma}\Sigma^{(-1)}(\tau)
\end{array} \eqno(4.10)$$

\noindent where Eqs.(3.6), (3.9),(3.10), (3.28) and the expression (3.34) for
$\Sigma^{(-1)}(\tau)$ have been used.  Eq.(4.10) shows that
$\Sigma^{(0)}(\tau)$ and $\Sigma^{(-1)}(\tau)$ differs by only the constant
$(\gamma -2)/\gamma$ indeed.  By comparing Eq.(4.7) with Eq.(4.10) and using
the relation (4.6) we see that the equality  
$$\left(\frac{\gamma}{2}-1\right) \ {_{1}F_{1}}(\alpha;\gamma;-\tau)-
\frac{\alpha}{\gamma}\tau \ {_{1}F_{1}}(\alpha +1;\gamma +1;-\tau)=
\frac{\gamma -2}{\gamma}e^{-\tau}\left[
\frac{\gamma}{2} \ {_{1}F_{1}}(\bar{\alpha};\gamma;\tau)+
\frac{\bar{\alpha}}{\gamma}\tau \  {_{1}F_{1}}(\bar{\alpha}+1;\gamma+1;\tau)
\right] \eqno(4.11)$$
\noindent is valid.  From this equality it may be verified that the real 
solutions in Eqs.(3.19) and (3.25) satisfy the relations
$$\Sigma^{(0)}_{\rm irreg}(\tau)=\Sigma^{(-1)}_{\rm irreg}(\tau) \ {\rm and} \
\Sigma^{(0)}_{\rm reg}(\tau) = \Sigma^{(-1)}_{\rm reg}(\tau) \eqno(4.12)$$
\noindent It is emphasized that in the verification of Eq.(4.12), the equality
(4.11) hence the relation (4.6) plays a key role.  Since the constants
$A_i^{(0)}, A_r^{(0)}, A_i^{(-1)}$ and $A_r^{(-1)}$ are determined through the
UV asymptotic form of $\Sigma^{(s)}_{\rm irreg}(\tau)$ and
$\Sigma^{(s)}_{\rm reg}(\tau) \ (s=0,-1)$ this means that the relation (4.6)
must keep to be valid also in the UV asymptotic region.  It is just this
requirement that determines the selections of the phase factors in the
asymptotic expansions (3.23) and (3.38).\\
\indent In summary, the linearized Schwinger-Dyson equation (3.1) of the
fermion self-energy in the theory with running gauge coupling constant
has two linearly independent real solutions which can be considered as the
the complex solution $\Sigma^{(0)}(\tau)$ or $\Sigma^{(-1)}(\tau)$ or the
pairs of real solutions ($\Sigma_{\rm irreg}(\tau)\equiv
\Sigma^{(0)}_{\rm irreg}(\tau)=\Sigma^{(-1)}_{\rm irreg}(\tau),
\Sigma_{\rm reg}(\tau)\equiv \Sigma^{(0)}_{\rm reg}(\tau)=
\Sigma^{(-1)}_{\rm reg}(\tau)$).  The physical solutions
should be some linear combination of $\Sigma_{\rm irreg}(\tau)$ and
$\Sigma_{\rm reg}(\tau)$ with the combination coefficients to being determined
by the IR and UV boundary conditions (2.23) and (1.10).  The application of
these physical solutions to problem of chiral symmetry breaking will be
discussed elsewhere. \\    \\
{\Large \bf Acknowledgements} \\
\indent This work is suppored partially by National Natural Science Foundation
of China and by Grant No.LWTZ-1298 of the Chinese Academy of Sciences. \\

\noindent {\Large \bf References} \\
${}^{1}$ K.Lane, Phys. Rev. {\bf D10},2605(1974); H.D.Politzer, Nucl.
         Phys. {\bf B117},397(1976). \\
${}^{2}$ P.I.Fomin, V.P.Gusynin, V.A.Miransky and Yu.A.Sitenko,
              Rev. Nuovo Cimento {\bf 6},1(1983). \\
${}^{3}$ K.Higashijima, Phys. Rev. {\bf D29},1228(1984). \\
${}^{4}$ V.A.Miransky, Phys. Lett. {\bf B165},401(1985).   \\
${}^{5}$ C.Leung, S.Love and W.Bardeen, Phys. Rev. Lett.
              {\bf 56},1230(1986); Nucl. Phys. {\bf B273},649(1986). \\
${}^{6}$ D.Atkinson and P.W.Johnson, Phys.Rev.{\bf D37},2290(1988);
              {\it ibid.} {\bf D37},2296(1988);
              D.Atkinson, P.W.Johnson and K.Stam, Phys. Rev.
              {\bf D37},2996(1988); D.Atkinson and P.W.Johnson, Phys. Rev.
              {\bf D41},1661(1990).  \\
${}^{7}$ H.J.Munczek and D.W.Mckay, Phys. Rev. {\bf D39},888(1989). \\
${}^{8}$ H.J.Munczek and D.W.Mckay, Phys. Rev. {\bf D42},3548(1990).  \\
${}^{9}$ D.Atkinson, H.de Groot and P.W.Johnson, Phys. Rev.
              {\bf D43},218(1991).    \\
${}^{10}$ A.G.Willliams, G.Krein and C.D.Roberts, Ann. Phys.
               {\bf 210},464(1991).      \\
${}^{11}$ V.A.Miransky, M.Tanabashi and K.Yamawaki, Phys. Lett.
               {\bf B221},177(1989).       \\
${}^{12}$ Higher Transcendental Function, edited by A.Erd{\'{e}}lyi 
          (McGraw-Hill,New York, 1953), Vol.1;
          Zhu-Xi Wang and Dun-Ren Guo, An Introduction to Special
               Functions, ( Science Press, Beijing, 1979 ). \\

\end{document}